\documentclass[pra,twocolumn,showpacs,superscriptaddress,preprintnumbers,amsmath,amssymb,aps]{revtex4-1}
\usepackage{mathrsfs}
\usepackage{amsmath}
\usepackage{amssymb}
\usepackage{graphicx}
\usepackage{dcolumn}
\usepackage{bm}
\usepackage{subfigure}
\usepackage{color}
\usepackage{ifpdf}
\usepackage{epstopdf}
\usepackage{verbatim}
\usepackage{bbm}
\usepackage{multirow}
\usepackage{booktabs}
\usepackage{color}
\usepackage{braket}
\usepackage[
pdfstartview=FitH,%
bookmarks=true,%
bookmarksopen=true,%
breaklinks=true,%
colorlinks=true,%
linkcolor=blue,anchorcolor=blue,%
citecolor=blue,filecolor=blue,%
menucolor=blue,pagecolor=blue,%
urlcolor=blue]{hyperref}
\usepackage{float}
\usepackage[toc,page,title,titletoc,header]{appendix}
\usepackage{ragged2e}
\usepackage{lipsum}

\usepackage{array}
\usepackage{booktabs}

\usepackage{graphicx}
\usepackage{epstopdf}

\usepackage{color}

\def\be{\begin{equation}}
	\def\ee{\end{equation}}
\def\bea{\begin{eqnarray}}
	\def\eea{\end{eqnarray}}

\begin{document}
\title{Scattering halos in strongly interacting Feshbach molecular Bose-Einstein condensates}
\author{Yuying Chen}
\affiliation {School of Physics and Electronics Engineering, Shanxi University, Taiyuan 030006, China}
\author{Zhengxi Zhang}
\affiliation{State Key Laboratory of Advanced Optical Communication Systems and Networks, School of Electronics, Peking University, Beijing 100871, China}
\author{Chi-Kin Lai}
\affiliation{State Key Laboratory of Advanced Optical Communication Systems and Networks, School of Electronics, Peking University, Beijing 100871, China}
\author{Yun Liang}
\affiliation{State Key Laboratory of Advanced Optical Communication Systems and Networks, School of Electronics, Peking University, Beijing 100871, China}
\author{Hongmian Shui}
\affiliation{State Key Laboratory of Advanced Optical Communication Systems and Networks, School of Electronics, Peking University, Beijing 100871, China}
\affiliation{Institute of Carbon-based Thin Film Electronics, Peking University, Shanxi, Taiyuan 030012, China}
\author{Haixiang Fu}
\affiliation{College of Information Science and Engineering, Huaqiao University, Xiamen, Fujian 361021, China}
\author{Fansu Wei}\email{wfs@pku.edu.cn}
\affiliation{State Key Laboratory of Advanced Optical Communication Systems and Networks, School of Electronics, Peking University, Beijing 100871, China}
\author{Xiaoji Zhou}\email{xjzhou@pku.edu.cn}
\affiliation{State Key Laboratory of Advanced Optical Communication Systems and Networks, School of Electronics, Peking University, Beijing 100871, China}
\affiliation{Institute of Carbon-based Thin Film Electronics, Peking University, Shanxi, Taiyuan 030012, China}
\date{\today}

	\begin{abstract}	
    We investigate the scattering halos resulting from collisions between discrete momentum components in the time-of-flight expansion of interaction-tunable $^6\rm Li_2$ molecular Bose-Einstein condensates. A key highlight of this study is the observation of the influence of interactions on the collisional scattering process. We measure the production of scattering halos at different interaction levels by varying the number of particles and the scattering length, and quantitatively assess the applicability of perturbation theory. To delve into a general theory of scattering halos, we introduce a scattering factor and obtain a universal relation between it and the halo ratio. Furthermore, we simulate the formation of scattering halos under non-perturbative conditions and analyze the discrepancies between simulation results and experiments through a return pulse experiment. This study enhances our understanding of the physical mechanisms underlying scattering processes in many-body systems and provides new perspectives for further theoretical research.

	\end{abstract}
 \maketitle

\section{Introduction}

    Among the fundamental interaction phenomena within many-body systems, collisional scattering stands out as a key process with abundant mechanisms \cite{RevModPhys.71.1, PhysRevLett.118.240402, Schweigler2017}. The advent of laser cooling has made it possible to explore quantum many-body effects in cold atomic and molecular systems \cite{Donley2002, Regal2003, PhysRevLett.92.120403, Greiner2003, doi:10.1126/science.1093280}, spawning numerous theoretical and experimental studies of low-energy collisions \cite{PhysRevLett.93.090404, PhysRevLett.70.414, PhysRevLett.94.200401}. The scattering halo, with its indispensable role in calibrating collision scattering rates, has garnered significant research interest as a key phenomenon\cite{PhysRevLett.84.5462, PhysRevA.73.033602, PhysRevD.78.042003, PhysRevResearch.6.023217}.

    The application of optical lattices to quantum gases greatly enriches the means of manipulation \cite{PhysRevLett.82.2022, Denschlag2002}. Due to their high controllability, they offer powerful and versatile experimental methods for exploring novel quantum states \cite{PhysRevLett.97.190406, PhysRevLett.121.265301,PhysRevLett.99.070401}, high-precision metrology \cite{GUO20222291} \cite{Dong_2021} \cite{Hu2018}, quantum computing \cite{PhysRevA.104.L060601}, and other quantum simulations \cite{Bloch2012, PhysRevLett.121.265301, PhysRevA.107.023303, Shui:23}. Optical lattice pulses can rapidly modulate the external states of particles to prepare multiple discrete momentum components \cite{PhysRevD.78.042003, PhysRevA.92.013616}. They are separated during the time-of-flight (TOF) expansion \cite{PhysRevLett.101.155303}, allowing the direct observation of transport, interference, and collisional scattering between them.

    In the cases of ultra-cold gases, several studies have been conducted on the collisional scattering halos \cite{PhysRevLett.93.173201, PhysRevLett.93.173202, PhysRevLett.84.5462, PhysRevLett.94.200401, PhysRevA.73.033602}, including researches on multi-particle correlations \cite{Burdick_2016, Thomas_2017}, quantum thermalization \cite{PhysRevLett.117.235303}, quantum entanglement \cite{Thomas2018, Shin2019}, and ghost imaging within the field of quantum optics \cite{Khakimov2016}. However, previous work has only studied halos of small particle number density under weak interactions \cite{Hofferberth2008, PhysRevLett.95.220403, PhysRevD.78.042003, PhysRevResearch.6.023217}. To date, there have been no experiments to study the collisional scattering processes under strong interactions.

    In this work, we investigate the collisional scattering halos between discrete momentum components of $^{6}\rm {Li_2}$ molecular Bose-Einstein condensates (mBECs) produced by one-dimension (1D) lattice pulses during the TOF process. Experimentally, we study the impact of interactions on the scattering halos \(N_{\text{sc}}\) over a wide range by varying the s-wave scattering length \(a_s\) and the number of total particles \(N_{\text{0}}\). The scattering length is directly modulated by a Feshbach magnetic field \cite{RevModPhys.82.1225, Kotochigova_2014, PhysRevLett.81.69, PhysRevLett.94.103201}, and the number of condensed particles is controlled by the loading time of the magneto-optical trap (MOT). Theoretically, we use perturbation theory to quantitatively explain the changes in the scattering halos and identify their validity in different interaction regions. A scattering factor \(\gamma\) is introduced to derive a universal relation with the halo ratio \(P_{\text{sc}}\) in the experiment. Consequently, for comparison with the experiments, we study the changes in scattering halos by numerical simulations under non-perturbative conditions and conduct a return pulse experiment to explain the discrepancies.

   This paper is organized as follows. In Sec.\,\ref{sec:experiment}, we describe our experimental implementation. In Sec.\,\ref{sec:theoretic}, the classical model of collision for scattering halos in 1D optical lattice with strong interaction strength is described. In Sec.\,\ref{Result}, we present the experimental results of scattering halos with different interactions. Finally, we give the conclusion in Sec.\,\ref{conclusion}.

\section{Experimental implementation}\label{sec:experiment}

\subsection{Experimental design} %\label{Comparison}
    
  The overall experimental design is shown in Fig.\,\ref{fig:f1}. We design 1D lattice pulses by the shortcut method(see Appendix \ref{Sup:A} for details) \cite{Zhou_2018,PhysRevLett.121.265301,PhysRevA.87.063638}, to load the mBEC from state $\ket{\psi_0}=\ket{p=0}$ into state $\ket{\psi}=\frac{1}{\sqrt 2}(\ket{p=+2\hbar k}+\ket{p=-2\hbar k})$, where $p$ is the momentum, $\hbar$ is the reduced Planck constant, $k = \pi/D$ is the wave vector of the optical lattice, $D$ is the lattice constant. After the state preparation, the lattice beams and trapping potential are turned off instantaneously, and the particles with different momentum modes that initially overlap each other in space will separate. The total time required for the separation of particles with different momenta $+2\hbar k$ and $-2\hbar k$ is denoted by $T$, in which elastic two-body collisions occur and produce a scattering halo between the two momentum peaks. The entire process is visualized through time-of-flight absorption imaging.
  
  The formation of scattering halos is closely related to the interaction strength. For a trapped condensate, the interaction strength is positively correlated with \textit{s}-wave scattering length $a_s$ and particle density $n$. With different $a_s$ and $n$, the scattering halos formed during TOF are also different. 
   \begin{figure}
    \includegraphics[width=0.25\textwidth]{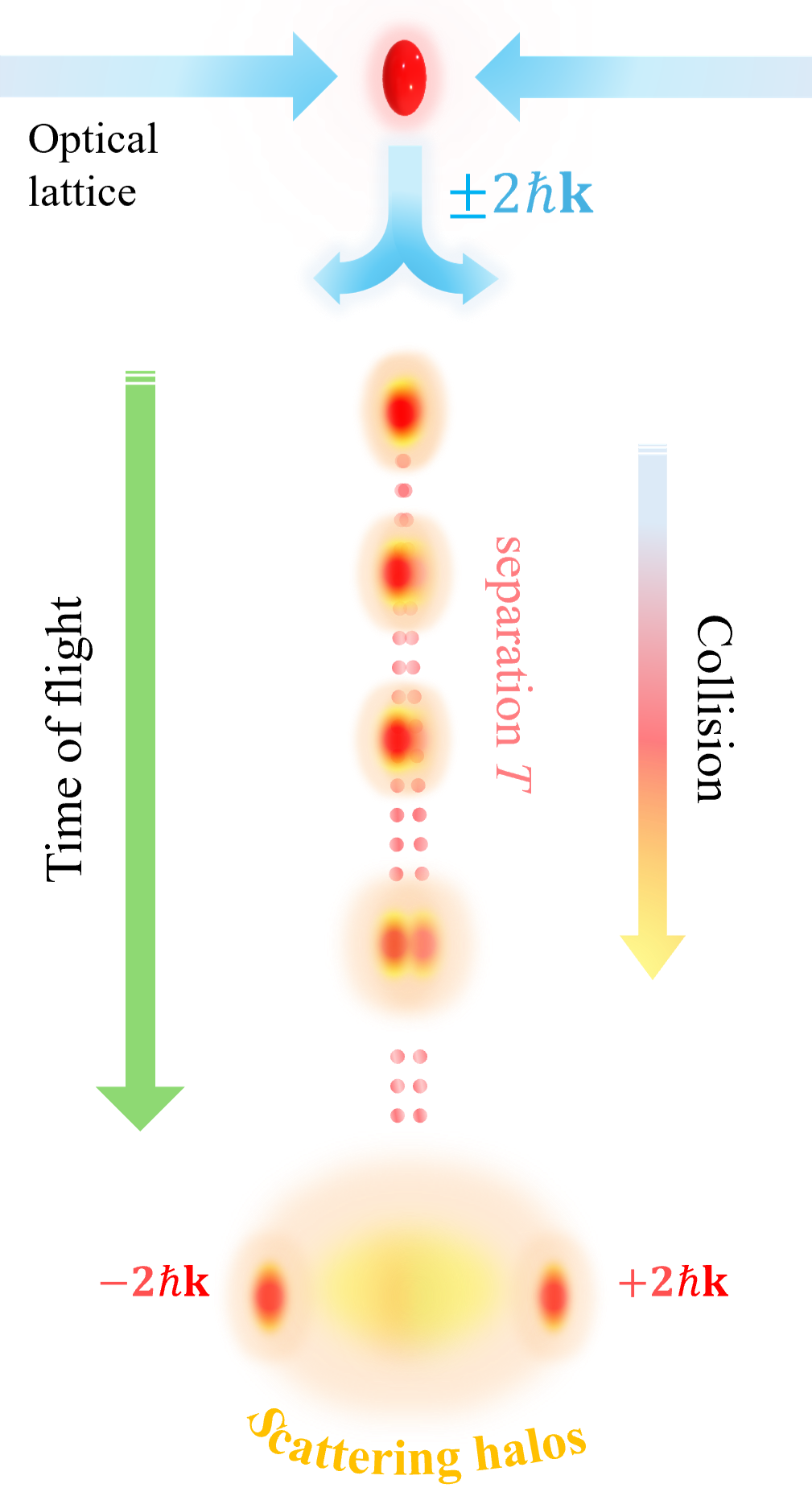}
    \caption{Scheme of the experimental design. The \(^{6}\rm Li_2\) molecular Bose-Einstein condensates (red particle clouds), pulsed by a 1D optical lattice (blue arrows), expand into free space after the lattice beams and trapping potential are instantaneously turned off. Due to the designed lattice pulses, a 50:50 proportion of \(\pm 2\hbar k\) momentum modes is formed. They pass through each other during the separation time $T$ in the TOF, leading to elastic two-body collisions. After this, the scattering halo (light yellow halos) will emerge between the two momentum peaks.}
	\label{fig:f1}
    \end{figure}

 \begin{figure*}[htp]
    \includegraphics[width=0.85\textwidth]{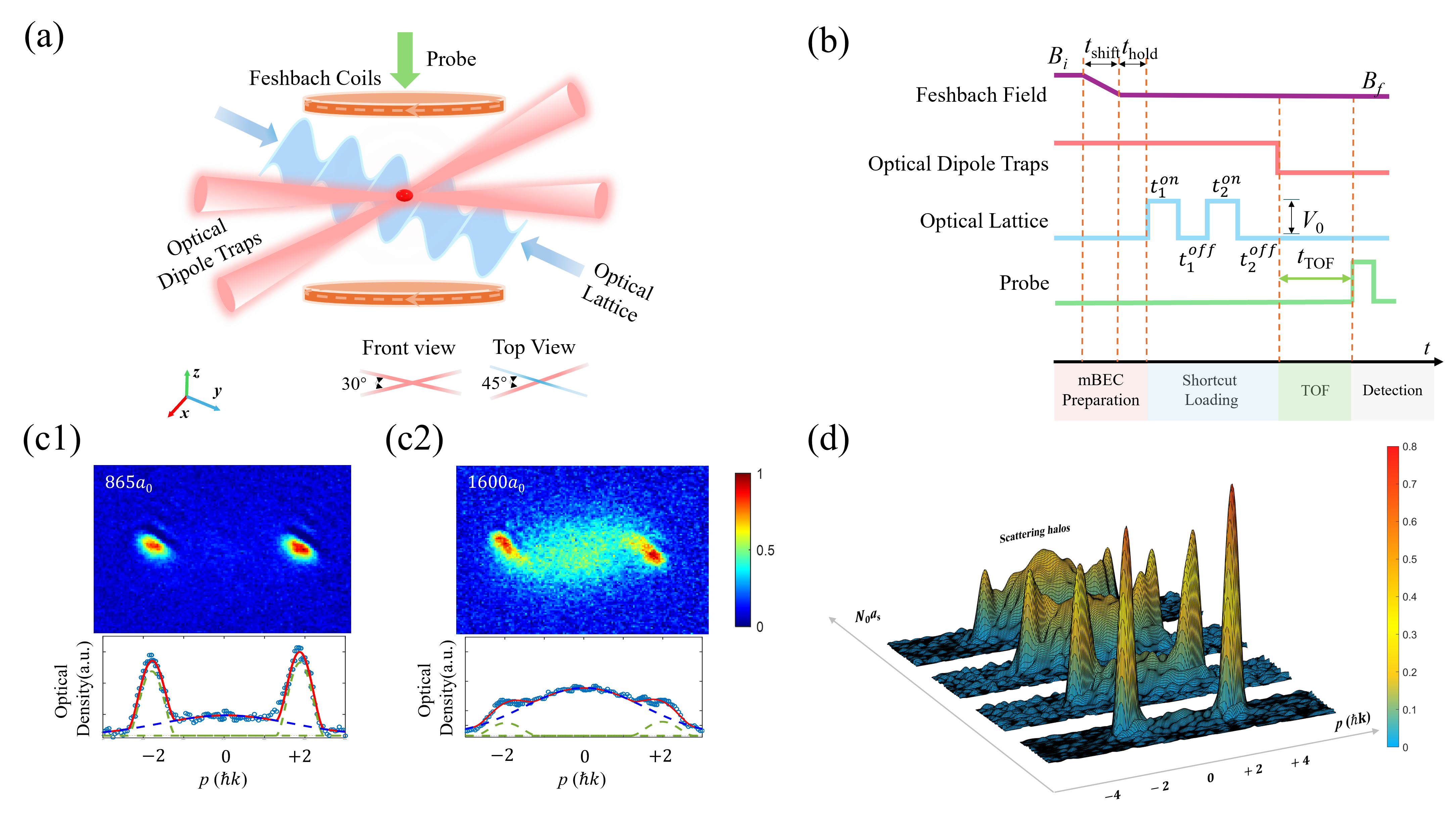}
    \caption{
    (a) Schematic of the experimental system. The brown circles in the $\textit{x}$-$\textit{y}$ plane represent the Feshbach magnetic coils. Feshbach mBECs are trapped in a pair of crossed optical dipole traps at $30^{\circ}$ to each other in the \textit{y}-\textit{z} plane (front view). The two blue arrows in the \textit{x}-\textit{y} plane represent the lattice beams, which form a $45^{\circ}$ angle with the optical trap (top view). The green arrow shows the imaging direction, which is perpendicular to the $\textit{x}$-$\textit{y}$ plane. (b) A typical experimental time sequence. The horizontal axis denotes the experimental time. The purple line and red line represent the Feshbach field and optical dipole trap, respectively. The Feshbach field is switched from $B_i$ to $B_f$ in $t_{\rm shift}$ and kept for a period of $t_{\rm hold}=100$ ms. The blue line represents the optical lattice. Several shortcut pulses \(t^{on(off)}_i\) are designed to load mBEC into the different momentum states $\pm 2\hbar k $ with lattice depth $V_0=10E_r$. The green line represents the probe beam, which is applied after the time-of-flight process with time $t_{\rm TOF}= 2\ \mathrm{m s}$. Different stages are marked at the bottom with varying colors. (c) corresponds to scattering patterns for different scattering strengths of 850$a_0$ (c1) and 1600$a_0$ (c2), with a fixed number of total particles \(N_{\text{0}}\) approximately equal to \(2 \times 10^4\). The corresponding curves below the images are obtained by integrating along the \textit{y} direction, showing the bi-modal fitting result for distinguishing between the condensed (the green dashed line) and non-condensed (the blue dashed line) part, and the light blue circles represent experimental results. The horizontal axis represents the momentum $p$. (d) A stereoscopic view of the two-dimensional particle momentum distribution observed through absorption imaging after a certain time of flight. The \textit{x}-axis of the horizontal plane coordinate represents different momentum states, and the \textit{y}-axis represents the interaction strength (quantified jointly by the number of mBEC \(N_{\text{0}}\) and the scattering length \(a_s\)). The color scale denotes the normalized density of particles.}
	\label{fig:f2}
   \end{figure*}

\subsection{Experimental setup and sequence} %\label{Comparison}
    
  We first prepare a two-state mixture of lithium atoms in their lowest hyperfine states \ \(|1\rangle\) = \(|F=1/2, m_F=1/2\rangle\) and \ \(|2\rangle\) = \(|F = 1/2, m_F = -1/2\rangle\). Subsequently, by performing evaporative cooling in the optical dipole trap at \(810\) Gauss of the Feshbach resonance and switching to the BEC side (ranging from \(655\) to \(766\) Gauss in our experiment), we can obtain $^6\rm{Li_2}$ mBECs with negligible thermal particles. The number of total particles \(N_{\text{0}}\) is adjusted by varying the loading time of the MOT, which has no impact on trap frequencies. The MOT loading time is varied from 1\,s to 8\,s corresponding to \(N_{\text{0}}\) ranging from \(1 \times 10^3\) to \(5 \times 10^4\). By controlling the Feshbach magnetic field, we can adjust the \textit{s}-wave scattering length between molecules, which is given by \(a_s = 0.6a_{12}\) \cite{PhysRevLett.93.090404}, where \(a_{12}\) is the \textit{s}-wave scattering length between atoms in states \(|1\rangle\) and \(|2\rangle\). In our experiment, the range of the intermolecular \textit{s}-wave scattering length \(a_s\) spans from \(865a_0\) to \(2623a_0\), where $a_0$ is the Bohr radius ($0.0529\ \mathrm{nm}$). The detailed experimental setup has been described in our previous work \cite{PhysRevA.109.043313}. Figure.\,\ref{fig:f2}(a) shows the schematic of the experimental setup. The mBEC is confined in a trapping potential formed by a pair of far-red-detuned lasers in a vertical plane with a $30^{\circ}$ to each other. The combined trapping frequencies are $(\omega_x,\omega_y,\omega_z) = 2\pi \times (39.5, 200, 200)$ Hz. The 1D optical lattice, whose lattice constant $D = \lambda/2 = 532\ \rm nm$, is composed of a $\lambda = 1064\ \rm nm$ laser beam and its reflected beam. The lattice beam intensity is controlled by an acousto-optic modulator (AOM) with feedback control, and another AOM is used for a synchronous on-off radio frequency (RF) switch. The characteristic lattice energy is $E_r = \hbar^2 \mathit{k}^2/2\mathit{m} $, where $k = \pi/D$ and $m$ is the mass of a lithium molecule ($^6\rm{Li_2}$).  

  The experimental time sequence is presented in Fig.\,\ref{fig:f2}(b). The horizontal axis indicates the experimental time. The first stage is the preparation of mBECs. The Feshbach field is switched from $B_i$ to $B_f$ in $t_{\rm shift}$ and kept for a period of $t_{\rm hold}=100$ ms. In the second stage, the mBEC is loaded into different momentum states via a series of shortcut lattice pulses. The lattice depth $V_0=10E_r$, and the on/off time of the two-pulse sequence is $\ ({{t} ^ {on} _ {1}, {t} ^ {off} _ {1}, {t} ^ {on} _ {2}, {t} ^ {off} _ {2}} \ )=\ ({12.3, 6.8, 7.1,18.2} )\ \rm \mu s$ (see Appendix \ref{Sup:A} for details). Absorption images are taken after the TOF process ($t_{\rm TOF}= 2\ \mathrm{m s}$) to measure the distribution of particles. Two typical images under different interactions (865$a_0$ and 1600$a_0$) are shown in Fig.\,\ref{fig:f2}(c). By integrating along the $y$-axis, we obtain the corresponding one-dimensional density distribution function of mBECs along the lattice, as shown in the bottom panel of Fig.\,\ref{fig:f2}(c). The total particle number \(N_{\text{0}}\) is obtained by integrating the 1D density distribution function. Then, we use a bimodal fitting method to determine the particle number of condensed parts (the green dashed line) and the thermal part (the blue dashed line). The number of scattering halos (\(N_{\text{sc}}\)) is equal to the latter. In this manner, we can quantitatively determine the scattering halo situation under different interaction conditions. 
  
  By comparison, we find that greater interactions lead to a more pronounced scattering halo in our experiment. Figure.\,\ref{fig:f2}(d) provides a stereoscopic perspective of the two-dimensional particle distribution across various interactions after a definite time of flight. We attempt to explore the impact of the scattering length $a_s$ and the initial particle number $N_{\rm 0}$ (directly related to density) on the formation of scattering halos by quantifying the number of scattering halos $N_{\rm sc}$ in the latter. 
   
\section{The perturbation theory of scattering halos}\label{sec:theoretic}

   To gain a deeper understanding of the experiment, we conduct a theoretical analysis of the variations in scattering halos under different interactions. This theoretical framework is primarily based on the first-order perturbation theory of the classical collision model, which has been mentioned in previous research \cite{PhysRevLett.84.5462, PhysRevLett.94.200401, PhysRevA.73.033602}. However, it mainly focused on scattering halo phenomena at smaller scattering lengths (where \(a_s\) is approximately \(10^2 a_0\)) and has not been specifically analyzed at larger scattering lengths.

   Under the perturbation theory, the number of particles scattered per unit time can be expressed as
   \begin{equation}
    \frac{\rm d}{\rm dt} N_{\text{sc}} = 2 \int d \vec{r} \sigma n_{+}(\vec{r},t) n_{-}(\vec{r},t) v
   \label{eq:formula(1)}
   \end{equation}
   \(\sigma\) represents the \textit{s}-wave scattering cross-section, which is related to the scattering length \(a_s\) and given by \(\sigma = 8\pi a_s^2\). \(n_+(\vec{r},t)\) and \(n_-(\vec{r},t)\) denote the densities of particles with momenta \(+2\hbar k\) and \(-2\hbar k\) at position \(\vec{r}\) and time \(t\), respectively. The relative velocity \(v\) between the two colliding bodies is determined by the optical lattice pulses with \(v = \frac{4\hbar k}{m}\). We use \(\Gamma\) to describe the two-body collision rate between particles, written as \(\Gamma = \sigma n_{+}(\vec{r},t) n_{-}(\vec{r},t) v\), representing the number of collisions occurring per unit time at position \(\vec{r}\) and time \(t\). 
   
   When the $a_s$ and $N_{\text{0}}$ are small enough, the density distribution \( n_+(\vec{r},t) \) and \( n_-(\vec{r},t) \) can be taken approximately as the distribution \( n_+(\vec{r}-\frac{1}{2}\vec{v}t,0) \) and \( n_-(\vec{r}+\frac{1}{2}\vec{v}t,0) \). Considering the initial distribution as Thomas-Fermi distribution and integrating Eq.\,(1) from $t = 0$ to $t = T$, we obtain the total number of particles in the scattering halo, as shown in Eq.\,\eqref{eq:formula(2)}. $T$ is the total time required for the separation of particles with momenta \( +2\hbar k \) and \( -2\hbar k \) (see Fig.\,\ref{fig:f1}). Then we can get the cumulative number of particles scattered as a result of two-body collisions during the TOF expansion:
   \begin{equation}
    N_{\text{sc}} = \left[\frac{5}{2} \epsilon^{1/3} \right]^2 \left(\frac{a_s N_{\text{0}}}{R}\right)^2
   \label{eq:formula(2)}
   \end{equation}
   $\epsilon$= 0.395 is the aspect ratio of mBECs in our experiment. The associated Thomas-Fermi radius $R$ is straightforwardly evaluated as
    \begin{equation}
    R = \left( \frac{15 N_{\text{0}} a_s \bar{a}^4}{\pi} \right)^{1/5} 
   \label{eq:formula(3)}
   \end{equation}
   where \(\bar{a} = [\hbar/(m \bar{\omega})]^{1/2}\) and \(\bar{\omega} = (\omega_{x} \omega_{y} \omega_{z})^{1/3}\). The relationship describes in Eq.\,\eqref{eq:formula(2)} can be further simplified to \(N_{\text{sc}} \propto (a_s N_{\text{0}})^{8/5}\) \cite{PhysRevResearch.2.013017} \cite{Chatelain_2020}.

   \section{Experimental results and analysis}\label{Result}
  
\begin{figure*}[htp]
    \includegraphics[width=0.85\textwidth]{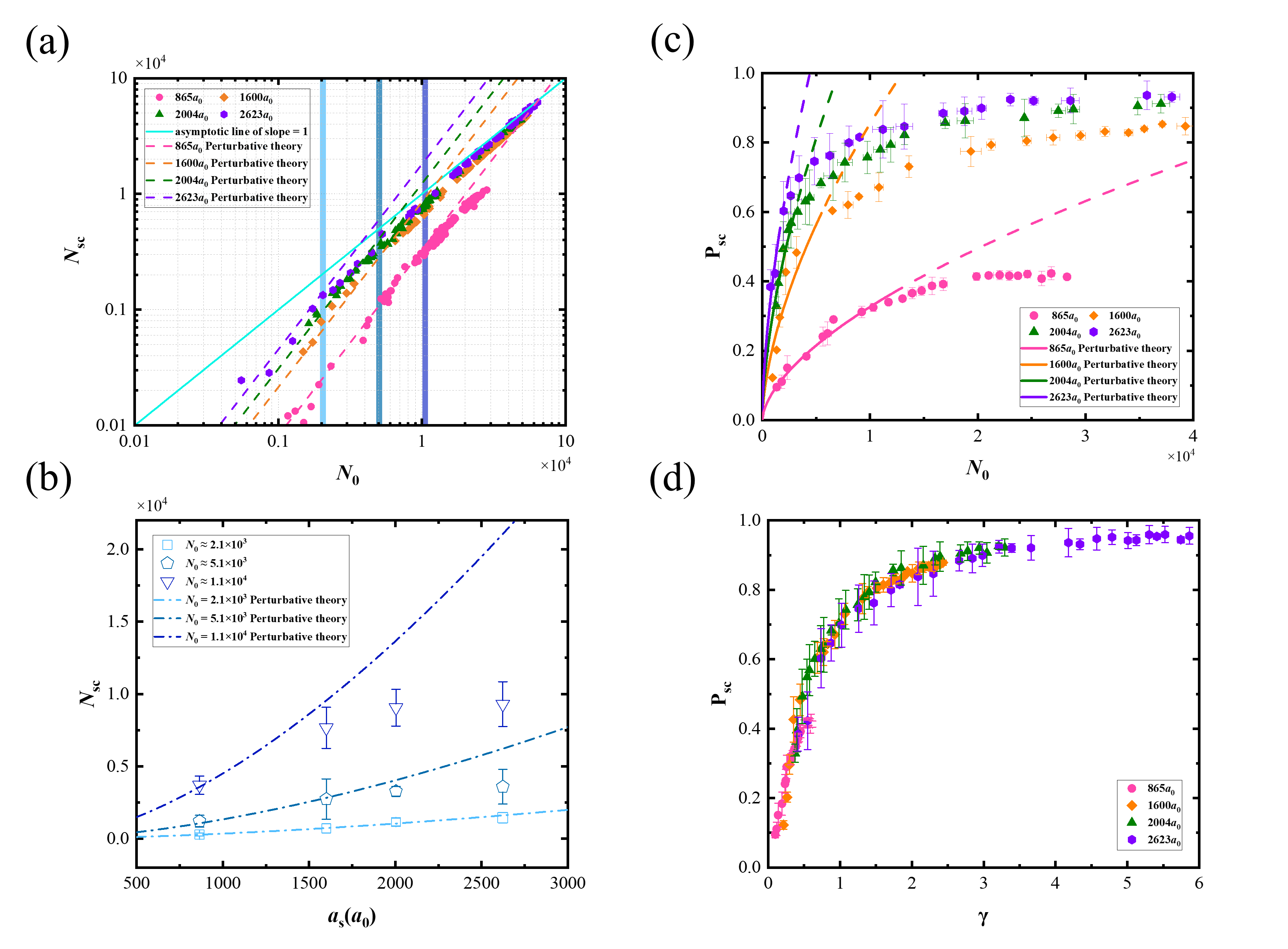}
    \caption{The effect of interactions on scattering halos. (a) The particle number of halo \(N_{\rm {sc}}\) with various total particle numbers \(N_{\rm 0}\) in a logarithmic coordinate system. The pink circles, yellow diamonds, green triangles, and purple hexagons represent the experimental data for scattering lengths \(a_s = 865a_0, 1600a_0, 2004a_0\), and \(2623a_0\), respectively. The corresponding colored dashed lines are the perturbation theoretical curves. The cyan solid line indicates the benchmark line with a slope of 1. Three vertical bands indicate different total particle numbers corresponding to \(N_{\rm {0}} \approx 2.1 \times 10^3\) (light blue), \(5.1 \times 10^3\) (blue), and \(1.1 \times 10^4\) (dark blue), with an uncertainty of \( 5\% \) for the particle number ranges. (b) The relationship between the number of scattering halos \(N_{\rm {sc}}\) and different scattering lengths \(a_s\) at various total particle numbers \(N_{\rm 0}\). The light blue squares, blue pentagons, and dark blue triangles mark the experimental data of $N_0\approx 2.1 \times 10^3$, \(5.1 \times 10^3\), \(1.1 \times 10^4\), and the dashed lines represent corresponding the perturbation theoretical curves. Error bars indicate the standard error of five measurements. (c) The relationship between the scattering halo ratio \(P_{\rm sc}\), which is the ratio of \(N_{\rm sc}\) to the total particle number \(N_{\rm 0}\), and the total particle number \(N_{\rm 0}\). The lines of different styles correspond to the perturbation theoretical curves for different scattering lengths \( a_s \). The experimental results are the average results of five measurements, and the corresponding error bars represent the uncertainty. (d) By analyzing the experimental data, we introduce the scattering factor \(\gamma\) and obtain a universal relationship between the scattering halo ratio \(P_{\rm sc}\) and the \(\gamma\). The corresponding error bars represent the uncertainty of \(P_{\rm sc}\).}
    \label{fig:f3}
\end{figure*}

   According to the experimental details in Sec.\,\ref{sec:experiment}, we can experimentally measure the number of scattering halos \(N_{\text{sc}}\) and the halo ratio \(P_{\text{sc}}=N_{\text{sc}}/N_{\text{0}}\), which is the ratio of scattering halos to total particles, under various \textit{s}-wave scattering lengths \(a_s\) and different initial particle numbers \(N_{\text{0}}\). Figure.\,\ref{fig:f3} shows experimental results with $a_s=$ \(865a_0\), \(1600a_0\), \(2004a_0\), \(2623a_0\), and $N_{\text{0}}$ ranging from \( 1\times 10^3\) to \( 5\times 10^4\). In Sec.\,\ref{Comparison}, we analyze the results of different interaction levels and compare them with the theoretical results of the perturbation theory to quantitatively assess its applicability. In Sec.\,\ref{Universal}, we introduce a scattering factor \(\gamma\) (as shown in Eq.\,\eqref{eq:formula(4)}), which indicates the likelihood of particles being scattered during the TOF process, to explore the general theory of scattering halos. Through calculating the relation between \(P_{sc}\) and \(\gamma\), we find a universal result as $P_{\text{sc}}=f(\gamma)$. We then simulate the formation process of scattering halos in the non-perturbative case (see in Sec.\,\ref{Simulation}), and the simulation results are compared with perturbation theoretical results and experimental data, with discrepancies analyzed through return pulse experiments.

   \subsection{The effect of interactions on the particle number of halos} \label{Comparison}

   Figure.\,\ref{fig:f3}(a) demonstrates the experimental results of the scattering halos \(N_{\text{sc}}\) varying with the particle number \(N_{\text{0}}\) at different scattering lengths \(a_s\) on a logarithmic scale. Parallel dashed lines of various colors represent the perturbation theoretical curves derived from Eq.\,\eqref{eq:formula(2)} for different values of \(a_s\), with the power-law relationship depicted as a slope of \(8/5\) after logarithmic transformation. The experimental results indicate that when \(a_s\) is fixed, \(N_{\text{sc}}\) does not maintain a linear relationship with \(N_{\text{0}}\) in a logarithmic plot. When \(N_{\text{0}}\) is relatively small, the experimental data align well with the perturbation theoretical curves. However, as \(N_{\text{0}}\) increases, results of different \(a_s\) begin to deviate from the theoretical curves. The slopes of the experimental data decrease from \(8/5\) to 1 progressively, while the data points of different scattering lengths gradually approach a single solid cyan line for sufficiently large \(N_{\text{0}}\). It indicates that when \(N_{\text{0}}\) are sufficiently large, almost all particles will be scattered due to collisions during the experiment, as \(N_{\text{sc}} = N_{\text{0}}\).
   
   To illustrate the relationship between $N_{\text{sc}}$ and \( a_s\), we select data sets of three different particle numbers ($ N_{\text{0}} \approx 2.1 \times 10^3 $, \( 5.1 \times 10^3 \), \( 1.1 \times 10^4 \), uncertainty =\( 5\% \)) for further analysis. As shown in Fig.\,\ref{fig:f3}(b), the blue hollow blocks with varying shades represent experimental data, and the corresponding dashed lines are the perturbation theoretical curves for the respective particle numbers. It can be seen that at a small particle number $ N_{\text{0}} \approx 2.1 \times 10^3 $, results for different \( a_s \) all match the light blue dashed lines, showing a good agreement with the perturbation theory. Whereas when particle numbers $ N_{\text{0}} \approx 5.1 \times 10^3 $ and \( 1.1 \times 10^4 \), the experimental data points deviate from the theoretical curves, from partial agreement to complete deviation, with an increase in scattering length. This suggests that the halo particle number versus scattering length relationship is consistent with the perturbative theory only at small scattering lengths and deviates significantly at larger ones. 
   
   Above all, by analyzing the halo particle numbers at different total particle numbers and scattering lengths, we observe the limitation of the perturbative theory in accurately quantifying changes in scattering halos with the increase of the interaction strength. The relationship between the halo particle number and interactions in the strongly interacting regime requires further theoretical corrections.

   %\begin{figure}
	%\includegraphics[width=0.45\textwidth]{figure4.png}
	%\caption{\,(a) The relationship between the scattering halo ratio \(P_{\rm sc}\), which is the ratio of \(N_{\rm sc}\) to the total particle number \(N_{\rm 0}\), and the total particle number \(N_{\rm 0}\). Pink circles, yellow diamonds, green triangles, and purple hexagons represent the experimental data for scattering lengths \(a_s = 865a_0, 1600a_0, 2004a_0\), and \(2623a_0\) respectively. The lines of different styles correspond to theoretical curves for different scattering lengths \( a_s \). The experimental points are the average results of five measurements, and the corresponding error bars represent the uncertainty. (b) By analyzing the experimental data, we introduce the scattering factor \(\gamma\) and obtain a universal relationship between the scattering halo ratio \(P_{\rm sc}\) and the \(\gamma\). The corresponding error bars represent the uncertainty of \(P_{\rm sc}\).}
	%\label{fig:f4}
%\end{figure}
   
   \subsection{Universal characterization of scattering halos} \label{Universal} 
  Due to the limitation of the perturbative theory in scattering halos, we attempt to experimentally generalize a universal relation between the halo particle number and interactions. Intuitively, we think it is more likely to uncover the universal relation by the ratio of the halo particle number to the total number of particles. Figure.\,\ref{fig:f3}(c) demonstrates the impact of different \( a_s \) and \( N_{\text{0}} \) on the halo ratios \( P_{\text{sc}} \), yielding similar conclusions to those of Fig.\,\ref{fig:f3}(a). The experimental data is processed by averaging, with vertical error bars indicating the uncertainty of $P_{\text{sc}}$, and horizontal error bars representing the standard deviation of $N_{\text{0}}$. Initially, when $N_{\text{0}}$ is low, the experimental results closely match the theoretical curves. However, as $N_{\text{0}}$ increases, the theoretical results exceed the experimental results. Notably, the $N_{\text{0}}$ threshold at which experimental results start to diverge from the theoretical curve differs with \( a_s \), suggesting that the formation of the scattering halo is jointly influenced by both \( N_{\text{0}} \) and \( a_s \).
  
  Therefore, to better access the relation between halo ratios \( P_{\text{sc}} \) and interactions, we introduce a factor \(\gamma\) that describes particle scattering, which is expressed as:
   \begin{equation}
   \gamma = \frac{N_{\text{sc}}^{\text{theory}}}{N_{\text{0}}}=\left[\frac{5}{2} \epsilon^{1/3} \right]^2 \frac{a_s^2 N_{\text{0}}}{R^2}
   \label{eq:formula(4)}
   \end{equation}
   \(N_{\text{sc}}^{\text{theory}}\) is the theoretical value of the number of scattering halos calculated based on Eq.\,\ref{eq:formula(2)}. In terms of the physical meaning of \(\gamma\), it can be considered in conjunction with Eq.\,\eqref{eq:formula(1)}, \(\gamma \sim a_s^2 N_{\text{0}} R^{-2} \sim \sigma n R \sim \Gamma T \), which corresponds to the scattering variation of particles after colliding with each other during the TOF expansion. 
   
   In light of \(\gamma\) being influenced by both the scattering length \(a_s\) and the initial particle number \(N_{\text{0}}\), we incorporate the \(\gamma\) to re-evaluate the changes in the scattering halo ratio \(P_{\text{sc}}\). As shown in Fig.\,\ref{fig:f3}(d), we find that under different scattering lengths \(a_s\), the scattering halo ratio \(P_{\text{sc}}\) tends towards a universal relation at different \(\gamma\), represented as \(P_{\text{sc}} = f(\gamma)\). Since the experimentally measured number of total particles \(N_{\text{0}}\) under different \(a_s\) have different maximum values, the corresponding measurement range for \(N_{\text{0}}\) also varies. This discrepancy is particularly evident in Fig.\,\ref{fig:f3}(d), as the newly introduced \(\gamma\) encompasses both \(N_{\text{0}}\) and \(a_s\). Nonetheless, the overlap between the data sets suggests that they are almost on the same trend line. To elucidate the functional form of this universal relationship more effectively, we present additional explanations and necessary refinements in Sec.\,\ref{Simulation}.

\begin{figure}
	\includegraphics[width=0.5\textwidth]{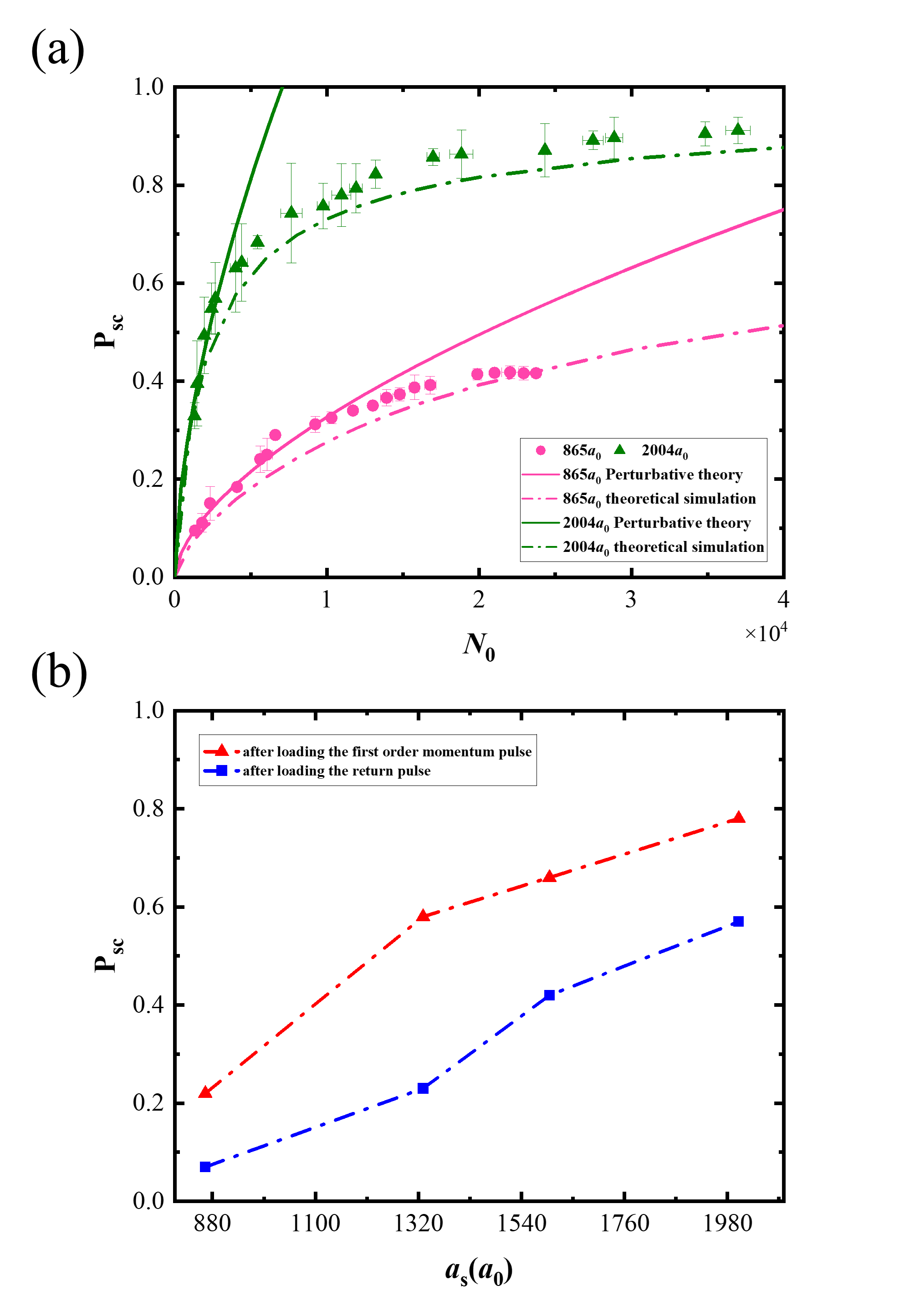}
	\caption{\,(a) Comparative analysis of numerical simulations and perturbation theory for the halo ratio \(P_{\text{sc}}\) as a function of the particle number \(N_{\text{0}}\). The corresponding scattering lengths are \(a_s = 865a_0\) (the pink circles) and \(2004a_0\) (the green triangles), with solid lines representing the perturbation theoretical curves and dashed lines representing the numerical simulation curves. (b) Analysis of the variation in the non-condensed halo ratio at different stages of pulse loading, with a fixed particle number of \(1.4 \times 10^4\), as the scattering length \(a_s\) changes. The experimental data after loading the first-order momentum lattice pulse is represented by red triangles with dashed lines, and the data after applying a return lattice pulse following the first-order momentum lattice pulse are represented by blue squares with dashed lines.}
	\label{fig:f5}
\end{figure}

 \subsection{Simulation for the non-perturbative case} \label{Simulation}

   To quantitatively describe the entire scattering process, we make corresponding corrections to the perturbation theory. Specifically, Eq.\,\eqref{eq:formula(2)} within this theory is primarily applicable when \(\gamma \ll 1\). However, considering that the scattering halo density \(n_{\text{sc}}\) increases during the collision process, we believe that the scattering halo also participates in the entire collision process. Therefore, Eq.\,\eqref{eq:formula(1)} should be rewritten as:
   \begin{equation}
   \begin{aligned}
   &\frac{\rm d}{\rm dt} N_{\text{sc}}  = \int d\vec{r} \ \sigma v [ 2 n_{+}(\vec{r}, t) n_{-}(\vec{r}, t)  \\
   &+ n_{+}(\vec{r}, t) n_{\text{sc}}(\vec{r}, t)+ n_{-}(\vec{r}, t) n_{\text{sc}}(\vec{r}, t) ]\label{eq:formula(5)}
   \end{aligned}
   \end{equation}
  
    As the number of scattering halo particles increases, scattering between the condensate particles and those scattering halo particles also becomes stronger. By taking this into account, we obtain:
   \begin{equation}
   \begin{aligned}
   &\frac{\rm d}{\rm dt} n_{+}(\vec{r}, t)  = -\sigma v [  n_{+}(\vec{r}, t) n_{-}(\vec{r}, t) + n_{+}(\vec{r}, t) n_{\text{sc}}(\vec{r}, t)] \\
   &\frac{\rm d}{\rm dt} n_{-}(\vec{r}, t)  = -\sigma v [  n_{-}(\vec{r}, t) n_{+}(\vec{r}, t) + n_{-}(\vec{r}, t) n_{\text{sc}}(\vec{r}, t)] \\
   &\frac{\rm d}{\rm dt} n_{\text{sc}}(\vec{r}, t)  = -\frac{\rm d}{\rm dt} n_{+}(\vec{r}, t) - \frac{\rm d}{\rm dt} n_{-}(\vec{r}, t) \label{eq:formula(6)}
   \end{aligned}
   \end{equation}

   By accounting for diffusion of particle motion and leveraging Eq.\,\eqref{eq:formula(5)} and Eq.\,\eqref{eq:formula(6)}, we simulate the scattering process for different particle numbers \(N_{\text{0}}\) at scattering lengths \(a_s\) of 865$a_0$ and 2004$a_0$. We analyze the relationship between the halo ratio \(P_{\text{sc}}\) and the particle number \(N_{\text{0}}\), comparing it with perturbation theory and experimental data. The parameters used in the simulation are all consistent with the experimental parameters. Meanwhile, we assume that the diffusion of the scattering halo is uniform in all directions. 
   
   As shown in Fig.\,\ref{fig:f5}(a), the solid lines mark the results of Eq.\,\eqref{eq:formula(2)}, while the dashed lines are the simulation results based on Eq.\,\eqref{eq:formula(5)}. The triangular and circular dots are the experimental results of 865\(a_0\) and 2004\(a_0\), respectively (the same in Fig.\,\ref{fig:f3}(c)). When \(\gamma\) is small, the experimental results are in good agreement with both the solid and dashed lines. Due to the decrease in the number of total particles during the TOF process, the simulation and the experimental results are both lower than the perturbation theoretical curves. Under the condition of the same particle number \(N_{\text{0}}\), the deviation between experimental and simulation results at 2004\(a_0\) is somewhat larger than at 865\(a_0\), which also reflects that the larger the value of \(\gamma\), the lower the applicability of Eq.\,\eqref{eq:formula(2)}. The reasons for this difference may be from the semiclassical nature of the simulation itself, experimental errors, and the deviation of the initial condensate ratio from 1, among others. Upon our analysis, the primary cause is likely the inter-particle scattering that takes place during the optical lattice pulses.

   To assess the loss of particles during the lattice pulse sequence, we utilize a return pulse technique to isolate the effects of the pulses \cite{SciPostPhys2022} \cite{10.21468/SciPostPhys.9.4.058}. By applying this designed lattice pulse sequence $\ ({{t}^{on}_{r1}, {t}^{off}_{r1}, {t}^{on}_{r2}, {t}^{off}_{r2}}\ )=\ ({19.3, 13.1, 12.5, 14.8})\ \rm \mu s$ to the state $\ket{\psi}=\frac{1}{\sqrt 2}(\ket{p=+2\hbar k}+\ket{p=-2\hbar k})$, the populations of higher momentum modes can be ``returned'' to the \(0\hbar k\) mode (see Appendix \ref{Sup:A} for details). Ideally, since the particles do not have different momentum components, there will be no significant collisional scattering during the TOF. Then the decrease in the number of total particles can be mainly attributed to inter-particle scattering during the lattice pulses. 

   Figure.\,\ref{fig:f5}(b) presents the result of \(P_{\text{sc}}\) with various scattering lengths \(a_s\) at different stages. The particle number is fixed at \(1.4 \times 10^4\). The red triangles mark the halo results after the loading sequence, and the blue squares mark those after the return sequence. The former contains the scattering halos generated during the loading pulses and TOF, and the latter contains the scattering halos generated during the loading pulses and return pulses. By comparing them in Fig.\,\ref{fig:f5}(b), we can draw two preliminary conclusions. Firstly, the proportion of scattering halos formed in the TOF stage is much larger than that produced in the return pulse stage. In other words, the TOF stage will contribute much more to the scattering halo than the pulse stage. Secondly, the larger the scattering length \(a_s\) is, the more scattering halos are generated during the pulses and become more non-negligible. Although the duration of the return pulse is much longer than the loading pulse duration in our experiment, the return pulse experiment indicates that the scattering halos observed in the TOF process should be considered as being formed during both the lattice loading stage and the TOF stage. This explains the deviation between the experimental and simulation results in Fig.\,\ref{fig:f5}(a).

\section{Conclusion}\label{conclusion}
    
   In this work, we focus on the impact of interactions on the formation of collisional scattering halos. The experiments reveal that \(N_{\text{sc}}\) aligns with the \(8/5\) power-law prediction of perturbation theory solely when the number of total particles is below a specific threshold. However, as \(a_s\) and \(N_{\text{0}}\) increase, the deviation of experimental results from perturbation theory becomes more pronounced. To quantitatively assess the applicability of perturbation theory, we introduce a scattering factor \(\gamma\) and determine its correlation with \(P_{\text{sc}}\). Our experiments uncover a universal relationship, indicating \(P_{\text{sc}}=f(\gamma)\), although the exact functional form still requires further determination, it offers a line of thought for delving into the general theory of scattering halos. We simulate the formation of scattering halos outside the perturbative regime and analyze the discrepancies between simulated and experimental results using return pulse experiments.

   In summary, this study not only measures the impact of interactions on collisional scattering halos experimentally but also provides new insights into the theoretical study of quantum many-body problems in strongly interacting systems.

\section*{Acknowledgements}
	The authors thank Zekai Chen and Chen Li for their helpful suggestions. This work is supported by the National Natural Science Foundation of China (Grants No. 92365208, No.11934002, and No. 11920101004), National Key Research and Development Program of China (Grants No. 2021YFA0718300 and No. 2021YFA1400900).

\appendix
    \addcontentsline{toc}{section}{Appendices}\markboth{APPENDICES}{}
    \begin{subappendices}

    \section{Shortcut Sequences}\label{Sup:A}

   The shortcut method, due to its robustness \cite{Zhou_2018,PhysRevLett.121.265301,PhysRevA.87.063638},  is utilized to prepare BEC into target states in an optical lattice. The basic idea of the shortcut method is to continually turn on and off the optical lattice to modulate the particle state, as shown in Fig.\,\ref{fig:f2}(b). In our experiment, the initial state \( |\psi_i\rangle \) is the ground state of mBEC in a harmonic trap. After several pulses, the final state of particles \( |\psi_f\rangle \) is written as:
   \begin{equation}
    |\psi_T\rangle = \prod_{i=1}^{n} \hat{U}_{\text{off}}(t^{\text{off}}_{i}) \hat{U}_{\text{on}}(t^{\text{on}}_{i})
    |\psi_i\rangle,
   \end{equation}
   where $n$ is the number of pulses, $\hat{U}_{\text{on}}$ and $\hat{U}_{\text{off}}$ are the evolution operators of the particle state when the optical lattice is on and off, respectively, and $t^{\text{on}}_{i}$ and $t^{\text{off}}_{i}$ are the evolution times of the $i$-th pulse.
    \begin{table}[htp]
    \caption{The pulse sequences used to prepare particles into the target state $\pm 2\hbar k $ of the 1D optical lattice with lattice depth $V_0=10E_r$.}
    \label{tab:initial_seq}
    \begin{tabular}{cccccccccc}
    \hline
    \specialrule{0em}{1pt}{1pt}
        {$V_0$}    & {$t^{\rm on}_1[\mu \rm s]$}    &{$\ \ t^{\rm off}_1\ \ $}    &{$\ \ t^{\rm on}_2\ \ $}    &{$\ \ t^{\rm off}_2\ \ $}      & {Theoretical fidelity}  \\
    \hline
    \specialrule{0em}{1pt}{1pt}
    {$10 E_r$} & {$12.3$}    & {$6.8$}    & {$7.1$}    & {$18.2$}    & {$99.9\%$} \\
    \hline
    \end{tabular}	
    \end{table}		
    
    Through designing the sequence, we can optimize the final state $\left | \psi_f \right \rangle$ to aimed state $\left | \psi_a \right \rangle$. The fidelity is defined by $|\left \langle \psi_f|\psi_a \right \rangle |^2$ to describe the loading efficiency. In the experiment, the optimized sequence has two pulses to load particles into the $\pm 2\hbar k $ momentum states of 1D optical lattices. The theoretical fidelity is calculated under non-interaction conditions. If interactions are considered, the loading fidelity of the shortcut method decreases somewhat. Although there is a drop, we can not observe condensates in other momentum states experimentally. Thus we conclude that this method is still valid within the strongly interacting regime.

    As shown in Fig.\,\ref{fig:f2}(c), the occupation numbers of momentum modes can be precisely controlled by lattice pulses. To study the impact of particle number loss during the lattice loading process on the scattering halo, we designed two additional return lattice pulses that can bring most particles back to the $0\hbar k$ mode before the time-of-flight. The specific pulses and theoretical fidelity are shown in Table \ref{tab:return_seq}.

    %As shown in Fig. \ref{fig:f2}(c), the occupation numbers of momentum modes are precisely controlled by lattice pulses, and two lattice pulses can transfer a BEC initially in the zero momentum mode to high momentum modes. To study the impact of particle number loss during the lattice loading process on the scattering halo results, we designed two additional return lattice pulses that can bring most particles back to the $0\hbar k $ mode before the time-of-flight. The pulse sequence is designed with specific pulse durations and intervals for a given lattice depth $V_0=10E_r$, aiming to maximize the overlap between the final wavefunction and the BEC wavefunction in the $0\hbar k $ mode. The specific pulses and theoretical fidelity are shown in Table \ref{tab:return_seq}.
    \begin{table}[htp]
    \caption{The shortcut sequences used to return the $0\hbar k $ mode in 1D optical lattices with lattice depth $V_0=10E_r$.}
    \label{tab:return_seq}
    \begin{tabular}{cccccccccc}
    \hline
    \specialrule{0em}{1pt}{1pt}
        {$V_0$}    & {$t^{\rm on}_{r_1}[\mu \rm s]$}    &{$\ \ t^{\rm off}_{r_1}\ \ $}    &{$\ \ t^{\rm on}_{r_2}\ \ $}    &{$\ \ t^{\rm off}_{r_2}\ \ $}      & {Theoretical fidelity}  \\
    \hline
    \specialrule{0em}{1pt}{1pt}
    {$10 E_r$} & {$19.3$}    & {$13.1$}    & {$12.5$}    & {$14.8$}    & {$99.3\%$} \\
    \hline
    \end{tabular}	
    \end{table}
\end{subappendices}

\bibliographystyle{apsrev4-2}
\bibliography{ref}

%\begin{refcontext}[sorting = none]
%\printbibliography
%\end{refcontext}
 
\end{document}